\documentclass[preprintnumbers,elsart]{revtex4}
\usepackage{amssymb}
\usepackage{amsmath}
\usepackage{graphicx}
\usepackage{dcolumn}
\usepackage[center]{subfigure}
\usepackage{float}
\usepackage{color}

\begin{document}

\title{Holding and transferring matter-wave solitons against gravity by
spin-orbit-coupling tweezers}
\author{Bin Liu$^{1,\S }$, Rongxuan Zhong$^{1,\S }$, Zhaopin Chen$^{2}$,
Xizhou Qin$^{1}$, Honghua Zhong$^{3}$}
\email{hhzhong115@163.com}
\author{Yongyao Li$^{1,2}$ and Boris A. Malomed$^{2,1}$}
\affiliation{$^{1}$School of Physics and Optoelectronic Engineering, Foshan University,
Foshan 528000, China \\
$^{2}$ Department of Physical Electronics, School of Electrical Engineering,
Faculty of Engineering, Tel Aviv University, P.O.B. 39040, Tel Aviv, Israel.%
\\
$^{3}$Institute of Mathematics and Physics, Central South University of
Forestry and Technology, Changsha 410004, China.\\
$^{\S }$ These two authors contributed equally to the work}

\begin{abstract}
We consider possibilities to grasp and drag one-dimensional solitons in
two-component Bose-Einstein condensates (BECs), under the action of gravity,
by tweezers induced by spatially confined spin-orbit (SO) coupling applied
to the BEC, with the help of focused laser illumination. Solitons of two
types are considered, semi-dipoles and mixed modes. We find critical values
of the gravity force, up to which the solitons may be held or transferred by
the tweezers. The dependence of the critical force on the magnitude and
spatial extension of the localized SO interaction, as well as on the
soliton's norm and speed (in the transfer regime), are systematically
studied by means of numerical methods, and analytically with the help of a
quasi-particle approximation for the soliton. In particular, a noteworthy
finding is that the critical gravity force increases with the increase of
the transfer speed (i.e., moving solitons are more robust than quiescent
ones). Nonstationary regimes are addressed too, by considering abrupt
application of gravity to solitons created in the weightless setting. In
that case, solitons feature damped shuttle motion, provided that the gravity
force does not exceed a dynamical critical value, which is smaller than its
static counterpart. The results may help to design gravimeters based on
ultracold atoms.
\end{abstract}

\maketitle

\section{Introduction}

Optical tweezers, which were first constructed in the classic works of
Ashkin \textit{et al.} \cite{Ashkin,Chu} for selecting and driving
microscopic dielectric particles by means of laser beams, are nowadays used
in various experiments aimed at trapping and transfer of atomic
Bose-Einstein condensates (BECs) \cite{Steuernagel2005}-\cite{Nieminen2014}.
Among these settings, is holding BEC\ under the action of gravity. The
interaction of ultracold condensates with the gravitational field is a
problem with applications to experiments in gravitational physics and
development of precise measurement techniques. In particular, atomic Raman
interferometry \cite{Fixler2007}-\cite{Xu2018} and Bloch oscillations \cite%
{Clade2005}-\cite{Poli2011} of ultracold atoms trapped in a vertical optical
lattice were used for precise measurement of the free-fall acceleration, as
well as its gradient, aimed at geological applications. A combination of
these techniques was reported to remarkably improve the accuracy of atomic
interferometry \cite{Fattori} and fundamental measurements of gravity \cite%
{Charriere}. Recently, a fountain gravimeter, which achieved a sub-$\mathrm{%
\mu }$G degree of accuracy, was fabricated on an atom chip \cite{Abend2016},
and a compact quantum gravimeter was realized in polarization-synthesized
spin-dependent optical lattices \cite{Yongguan2018}-\cite{Robens2018}.
Dynamics of BEC under the action of microgravity has also been a subject of
many experiments \cite{microgravity}-\cite{microgravity2}.

The self-attractive intrinsic nonlinearity of BEC, induced by inter-atomic
collisions, which is accurately modeled by the Gross-Pitaevskii equation
(GPE) \cite{GP}, gives rise to self-trapped states, in the form of solitons
\cite{recent-review} and breathers \cite{Strathclyde}. As concerns the
above-mentioned topics, relevant aspects are manipulations of matter-wave
solitons with the help of tweezers \cite{Carpentier2008}, motion of solitons
\cite{Holtzmann}-\cite{Rosanov2} and breathers \cite{Alodjants,Rosanov3}
driven by gravity, and Bloch oscillations of gap solitons in optical
lattices under the action of a constant (gravity) force \cite{Salerno2008}.
In this context, it is worthy to mention great diversity in the design of
matter-wave interferometers offered by matter-wave solitons~\cite{Rosanov4}-%
\cite{splitter}.

However, the applications of optical tools to BEC are hampered by effects of
heating the condensate by laser beams which are used for the creation of
tweezers \cite{heating1}-\cite{heating3} and optical lattices \cite%
{OL-heating1}-\cite{OL-heating7}. In this work, we aim to elaborate a
tweezers scheme for matter-wave solitons based on spatially localized
spin-orbit (SO) coupling applied to BEC, by means of properly focused laser
illumination. One of incentives for the introduction of the scheme is the
fact that the SO interaction may be induced in the condensate\ without
heating it \cite{no-heating1}-\cite{no-heating4}.

\begin{figure}[h]
{\includegraphics[width=0.25\columnwidth]{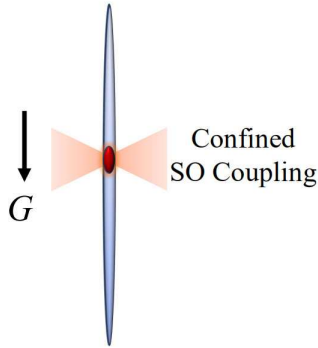}}
\caption{(Color online) The sketch of the setting: a quasi-1D matter-wave
soliton, designated by the dark red color, is created in the vertical
cigar-shaped potential trap, while horizontal beams, which induce the
confined SO coupling, hold the soliton at a particular vertical position. $%
{\protect\large G}$ represents the gravity force.}
\label{Sketch}
\end{figure}

The configuration is schematically shown in Fig. \ref{Sketch}, which assumes
that the soliton is placed in a narrow vertical cigar-shaped potential trap
and illuminated by a horizontal Raman laser beam inducing the SO coupling
and focused on a segment of the vertical \textquotedblleft cigar". In fact,
the focusing does not need to be very tight. Indeed, for a typical
longitudinal extension of solitons, which is, normally, several microns \cite%
{Randy}, the Raman-beam illumination should be confined on a comparable
segment, which is, thus, much larger than the diffraction limit for the
focusing, $\simeq \lambda /2$, as determined by the characteristic beam's
wavelength, $\lambda \simeq 1$ $\mathrm{\mu }$m \cite{Spielman,Galitski}.

The SO coupling in BECs, which emulates the eponymous effect in physics of
semiconductors, has been an active research subject in recent years \cite%
{Spielman}-\cite{Yongping2016}. In the combination with the usual
collision-induced nonlinearity, the SO coupling was theoretically shown to
support various species of one-dimensional (1D) solitons \cite{1D-SO-sol-1}-%
\cite{SKA0}. A surprising prediction is that the interplay of the linear SO
interaction and cubic self-attraction leads to stabilization of
multidimensional solitons in free space, which are completely unstable in
the absence of the SO terms \cite{Sakaguchi2014}-\cite{Pangw2018}.

As concerns the application of the SO coupling to a spatially confined
region, such as in the configuration sketched in Fig. \ref{Sketch}, it was
predicted that 1D solitons and their bound states may be maintained by means
of localized SO interaction \cite{1Dlocalized,Bin2019}. Further, it was
recently demonstrated that spatially confined SO coupling can be used to
maintain stable 2D solitons \cite{Yongyao2018}. A related possibility is the
use of effectively 1D and 2D (low-dimensional) SO coupling for the
stabilization of, respectively, 2D and 3D solitons (in the full dimension),
including a vortex component \cite{PRR}.

The subject of the present work is the stability and dynamics of quasi-1D
solitons held or transported by localized SO coupling against the action of
gravity, as suggested by Fig. \ref{Sketch}. The consideration is based on a
combination of systematic numerical simulations and the use of an analytical
approximation. The results may help to develop a new type of cold-atom
gravimeter and, more generally, new techniques for precise measurements.

The remainder of the paper is structured as follows. The setting is
introduced in Section II, which is followed by presentation of numerical and
analytical results for stationary solitons, obtained in both quiescent and
moving reference frames (the latter one pertaining to the transport of
solitons), in Section III. Motion (shuttle oscillations) of solitons is
numerically studied in Section IV, and the paper is concluded by Section V.

\section{The model}

The setting sketched in Fig. \ref{Sketch}, with the spatially confined SO
coupling of the Rashba type \cite{Rashba}, which can be created by the laser
beam \cite{Campbell2011}, is represented by the following system of coupled
GPEs for a pseudo-spinor (two-component) mean-field wave function, $\left(
\psi _{+},\psi _{-}\right) $, written in the scaled form:
\begin{eqnarray}
&&i\partial _{t}\psi _{+}=-{\frac{1}{2}}\partial _{xx}\psi _{+}+\Lambda
(x)\partial _{x}\psi _{-}-(|\psi _{+}|^{2}+\gamma |\psi _{-}|^{2})\psi _{+}+{%
\frac{1}{2}}\frac{d{\Lambda }}{dx}\psi _{-}-Gx\psi _{+},  \notag \\
&&i\partial _{t}\psi _{-}=-{\frac{1}{2}}\partial _{xx}\psi _{-}-\Lambda
(x)\partial _{x}\psi _{+}-(|\psi _{-}|^{2}+\gamma |\psi _{+}|^{2})\psi _{-}-{%
\frac{1}{2}}\frac{d{\Lambda }}{dx}\psi _{+}-Gx\psi _{-},  \label{Basiceq}
\end{eqnarray}%
where the strength of the attractive self-interaction in each component is
normalized to be $1$, $\gamma $ is the relative strength of the
cross-interaction between the components, and $G$ is the gravity force. In
the experiment, values of $G$ may be controlled by means of the angle
between the direction of the quasi-1D trap holding the condensate and the
vertical direction, as well as by running the experiment in microgravity
settings \cite{microgravity,microgravity2}. Because the natural transverse
structure of laser beams is Gaussian, we adopt the corresponding shape of
the spatial localization of the SO-coupling coefficient, cf. Refs. \cite%
{1Dlocalized} and \cite{Yongyao2018}:
\begin{equation}
\Lambda (x)=\Lambda _{0}\exp (-x^{2}/L^{2}),  \label{Lambda}
\end{equation}%
where $L$ defines the confinement size, while amplitude $\Lambda _{0}$ may
be fixed to $1$ by means of additional rescaling. Nevertheless, it is
convenient to keep $\Lambda _{0}$ as a free parameter, to explore dependence
of the predicted effects on the strength of the SO interaction.

Terms $\sim d\Lambda /dx$ in Eq. (\ref{Basiceq}) appear when the GPE system
is derived from the respective Hamiltonian,%
\begin{gather}
H=\int_{-\infty }^{+\infty }\frac{1}{2}\left[ \left( \left\vert \partial
_{x}\psi _{+}\right\vert ^{2}+\left\vert \partial _{x}\psi _{-}\right\vert
^{2}-\left\vert \psi _{+}\right\vert ^{4}-\left\vert \psi _{-}\right\vert
^{4}\right) -\gamma \left\vert \psi _{+}\right\vert ^{2}\left\vert \psi
_{-}\right\vert ^{2}\right.  \notag \\
+\frac{1}{2}\Lambda (x)\left( \psi _{+}^{\ast }\partial _{x}\psi _{-}-\psi
_{-}^{\ast }\partial _{x}\psi _{+}+\psi _{+}\partial _{x}\psi _{-}^{\ast
}-\psi _{-}\partial _{x}\psi _{+}^{\ast }\right)  \notag \\
\left. -Gx\left( \left\vert \psi _{+}\right\vert ^{2}+\left\vert \psi
_{-}\right\vert ^{2}\right) \right] dx.  \label{H}
\end{gather}%
Solutions are characterized by their total norm, which is a dynamical
invariant of the system,
\begin{equation}
N=\int_{-\infty }^{+\infty }dx\left( |\psi _{+}|^{2}+|\psi _{-}|^{2}\right) .
\label{N}
\end{equation}

Moving tweezers, which can transfer trapped solitons, are defined as
profiles (\ref{Lambda}) moving at velocity $c$,
\begin{equation}
\Lambda (x)=\Lambda _{0}\exp \left[ -\left( x-ct\right) ^{2}/L^{2}\right] .
\label{MovingSOC}
\end{equation}%
In particular, we aim to find a critical value of velocity, up to which the
trapped solitons may be stably transferred. This is a nontrivial issue
because Eq. (\ref{Basiceq}) is not Galilean invariant. To address it, we
follow Ref. \cite{Sakaguchi2014} and rewrite Eqs. (\ref{Basiceq}) in the
moving reference frame, with coordinate $x^{\prime }=x-ct$ and transformed
wave function, $\psi _{{\LARGE \pm }}\left( x,t\right) =\psi _{{\LARGE \pm }%
}^{\prime }\left( x^{\prime },t\right) \exp \left( icx^{\prime
}+ic^{2}t/2+iGct^{2}/2\right) $:
\begin{eqnarray}
&&i\partial _{t}\psi _{+}^{\prime }=-{\frac{1}{2}}\partial _{x^{\prime
}x^{\prime }}\psi _{+}^{\prime }+\Lambda (x^{\prime })\partial _{x^{\prime
}}\psi _{-}^{\prime }-(|\psi _{+}^{\prime }|^{2}+\gamma |\psi _{-}^{\prime
}|^{2})\psi _{+}^{\prime }+\left[ \frac{1}{2}{\frac{d\Lambda (x^{\prime })}{%
dx^{\prime }}+ic\Lambda }\left( x^{\prime }\right) \right] \psi _{-}^{\prime
}-Gx^{\prime }\psi _{+}^{\prime },  \notag \\
&&i\partial _{t}\psi _{-}^{\prime }=-{\frac{1}{2}}\partial _{x^{\prime
}x^{\prime }}\psi _{-}^{\prime }-\Lambda (x^{\prime })\partial _{x^{\prime
}}\psi _{+}^{\prime }-(|\psi _{-}^{\prime }|^{2}+\gamma |\psi _{+}^{\prime
}|^{2})\psi _{-}^{\prime }-\left[ \frac{1}{2}{\frac{d\Lambda (x^{\prime })}{%
dx^{\prime }}+ic\Lambda }\left( x^{\prime }\right) \right] \psi _{+}^{\prime
}-Gx^{\prime }\psi _{-}^{\prime }.  \label{eq3}
\end{eqnarray}%
Thus, the underlying equations indeed change in the moving frame. Note that
another transformation, $\psi _{\pm }^{\prime }\rightarrow \left( \psi _{\pm
}^{\prime }\right) ^{\ast }$, $t\rightarrow -t$, casts Eq. (\ref{eq3}) into
itself, with $c$ replaced by $-c$, thus making positive and negative values
of $c$ mutually equivalent [see also Fig. \ref{moving}(d) below].

The linear-mixing terms $\sim ic\Lambda $, which are generated by the
transformation in Eq. (\ref{eq3}), strongly affect the structure of
solitons, but, generally, do not make them unstable, as shown below.

\section{Stationary solutions}

\subsection{Stationary states in the quiescent reference frame}

\subsubsection{Numerical results}

In the 2D system, two types of fundamental solitons exist under the action
of the SO coupling. One is built as a composite state including a
fundamental soliton in component $\psi _{+}$ (one with zero topological
charge, $S_{+}=0$), and a solitary vortex (with charge $S_{-}=1$) in
component $\psi _{-}$. States of this type are classified as semivortex (SV)
solitons \cite{Sakaguchi2014} (similar composite modes, found in a model
with repulsive nonlinearity, were called \textquotedblleft half vortices"
\cite{BB2012}). Due to the symmetry of the SO coupling, SVs\ with the
\textit{opposite chirality}, defined by the vorticity set $%
(S_{+}=-1,S_{-}=0) $, exist too. The other type of 2D fundamental solitons
is the mixed-mode (MM) soliton, which is built as a superpositions of states
with vorticities $(0,1)$ and $(-1,0)$ in the two components.

A 1D counterpart of the SV soliton is a \textit{semidipole} (SD) composite
state, which is composed of spatially even and odd modes in its two
components, and can be found numerically starting from the following input:%
\begin{equation}
\psi _{+}=A_{1}\exp (-x^{2}/W^{2}),~\psi _{-}=A_{2}\left( x/W\right) \exp
(-x^{2}/W^{2}),  \label{SD}
\end{equation}%
with width $W$ and real amplitudes $A_{1,2}$. A 1D version of MM solitons is
built as a superposition of even and odd modes in both components \cite%
{Rongxuan2018}. Accordingly, the numerical search for such solutions may be
initiated by input
\begin{equation}
\psi _{\pm }=\left[ A_{1}\pm A_{2}\left( x/W\right) \right] \exp
(-x^{2}/W^{2}).  \label{MM}
\end{equation}%
The form of these is suggested by their 2D counterparts used in Ref. \cite%
{Sakaguchi2014} for constructing solitons of the SV and MM types.

In the numerical analysis, we chiefly focus on the Manakov's nonlinearity,
with $\gamma =1$ in Eq. (\ref{Basiceq}) \cite{Manakov} (although other
values of $\gamma $ are considered too). This case is relevant for
experimental realization of the SO coupling in binary mixtures of two
different hyperfine atomic states \cite{Galitski}-\cite{Yongping2016}, for
which scattering lengths of collisions between atoms belonging to the same
or different atomic states are nearly equal. The interplay of the SO
coupling and nonlinearity of the Manakov's type in the 2D system features
special properties. In particular, the SD and MM solitons are included, in
this case, in a broader family of 2D solitons with equal energies and equal
chemical potentials for a fixed soliton's norm (\ref{N}) \cite{Sakaguchi2014}%
. A systematic numerical analysis demonstrates below that states maintained
by the SO trap (\ref{Lambda}) are especially robust against the effect of
gravity in the system with the Manakov's nonlinearity.

Stationary solutions of Eq. (\ref{Basiceq}) for solitons can be obtained as
follows.

(i) Use the imaginary-time method (ITM) \cite{Chiofalo2000,JKyang2008} to
produce a stationary solution at $G=0$. Because the SD and MM solitons
represent mutually degenerate ground states in the Manakov's case, either of
them can be generated by the ITM with inputs chosen as per Eqs. (\ref{SD})
and (\ref{MM}), respectively (at $\gamma <1$ and $\gamma >1$, the ground
state is represented by solitons of the SD and MM types, respectively \cite%
{Sakaguchi2014}). Due to the SO coupling between the component of the
soliton, they share a common value of the chemical potential $\mu $, taking
the form of
\begin{equation}
\psi _{\pm }(x,t)=\phi _{\pm }(x)e^{-i \mu t}.  \label{stationary}
\end{equation}%
The chemical potential of the ITM-produced solutions can be accurately
identified by the substitution of the solutions in Eq. (\ref{Basiceq}) and
looking at the central point ($x=0$).

(ii) Use the wave function and chemical potential of the solution with $G=0$
as an initial guess, and then apply the squared-operator method (SOM) \cite%
{JKyang2007}, to produce a stationary solution with $G^{(1)}=\delta G$,
where $\delta G$ is a small value.

(iii) Using SOM, continue to build solutions, increasing the gravity force
by small steps, $G^{(n-1)}\rightarrow G^{(n)}=G^{(n-1)}+\delta G$.

In the case of $G>0$, the formal ground state (the one with the minimal
energy) corresponds to the free fall of the wave packet towards $%
x\rightarrow \infty $. For this reason, ITM, which seeks for the system's
ground state, is irrelevant for constructing solitons at $G>0$, and it was
necessary to use SOM at stages (ii) and (iii).

\begin{figure}[th]
{\includegraphics[width=0.8\columnwidth]{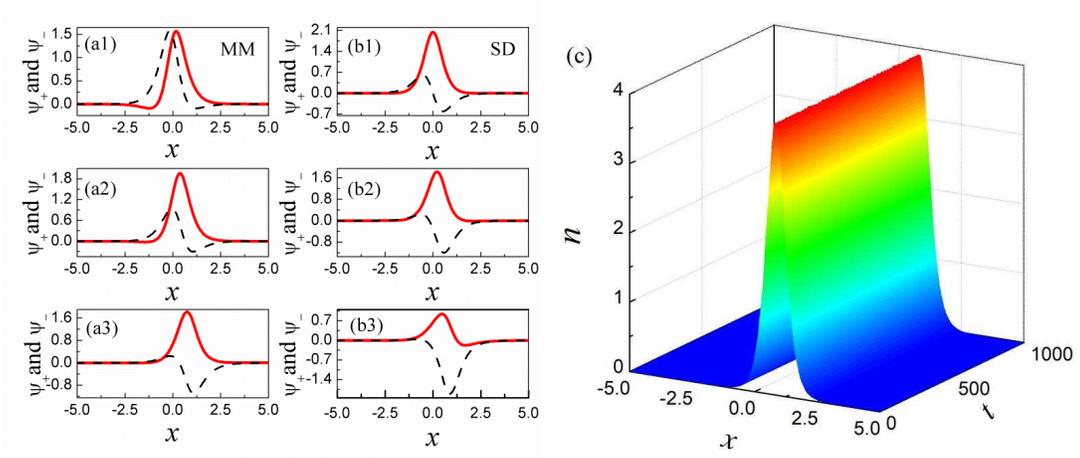}}
\caption{Continuous and dashed lines depict two components of stationary
wave functions in stable solitons of the MM and SD types [the left column,
(a1)-(a3), and the right one, (b1)-(b3), respectively] for different values
of the gravity force: $G=0$ (a1,b1), $0.12$ (a2,b2), and $0.24$ (a3,b3). The
solutions are obtained with a fixed chemical potential, $\protect\mu %
=-2.5231 $, starting from the MM and SD solitons with norm $N=4$ [see Eq. (%
\protect\ref{N})] at $G=0$. Solid and dashed curves in these panels
represent profiles of the $\protect\psi _{+}$ and $\protect\psi _{-}$
components, respectively. (c) Stability of the solitons (of the MM type, as
concerns this example) in direct simulations, for $G=0.24$. The soliton was
initially perturbed by random noise, added at the $3\%$ amplitude level. The
evolution is shown by the spatiotemporal distribution of the total density, $%
n(x,t)=|\protect\psi _{+}(x,t)|^{2}+|\protect\psi _{-}(x,t)|^{2}$. In this
figure, parameters of Eqs. (\protect\ref{Basiceq}) and (\protect\ref{Lambda}%
) are fixed as $\protect\gamma =1$ and $L=2$. }
\label{StableSDMM}
\end{figure}

Figures \ref{StableSDMM}(a,b) display typical examples of stable MM and SD
solitons, obtained by means of this procedure for different values of $G$
and fixed $\mu $. Stability of these solitons is illustrated by direct
simulations displayed in panel (c). The results show that the shape of MM
solitons quickly tends to become close to the SD type with the increase of $%
G $, therefore the consideration is focused below, chiefly, on the SD modes.

\begin{figure}[h]
{\includegraphics[width=0.8\columnwidth]{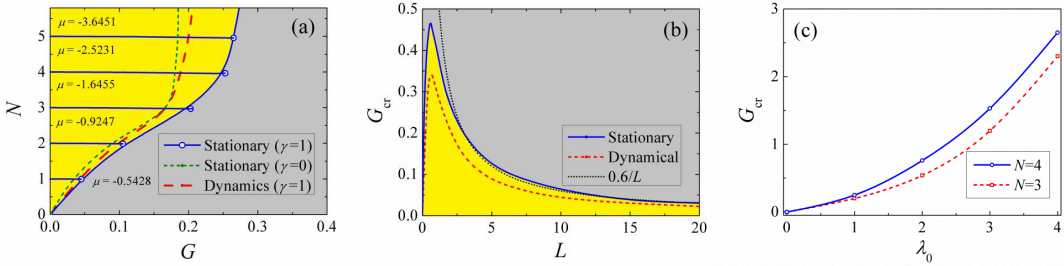}}
\caption{(a) Stability and instability areas (yellow and gray ones,
respectively) of stationary states in the $\left( N,G\right) $ plane for $%
\protect\gamma =1$. Here, the source soliton used at $G=0$ is one of the SD
type, and branches with $\protect\mu =-0.5428$, $-0.9247$, $-1.6455$, $%
-2.5231$, $-3.6451$ represent solution families starting with norms $%
N=1,2,3,4,5$ at $G=0$, respectively. The green dashed curve is the stability
boundary for $\protect\gamma =0$. Here and in panel (b), the red
dashed-dotted curve is the stability boundary of the shuttle-motion regime,
which is considered below. The SO confinement size is fixed in this panel as
$L=2$, see Eq. (\protect\ref{Lambda}). (b) The stability and instability
areas of stationary states in the $\left( G,L\right) $ plane, for fixed norm
$N=5$. The dotted line is the fitting curve, which is $G_{\mathrm{cr}}=0.6/L$%
. (c) $G_{\mathrm{cr}}$ as a function of $\Lambda _{0}$ for $N=4$ (solid
curve) and $N=3$ (dashed curve) for $(L,\protect\gamma )=(2,1)$.}
\label{Gcr}
\end{figure}

\begin{figure}[h]
{\includegraphics[width=0.8\columnwidth]{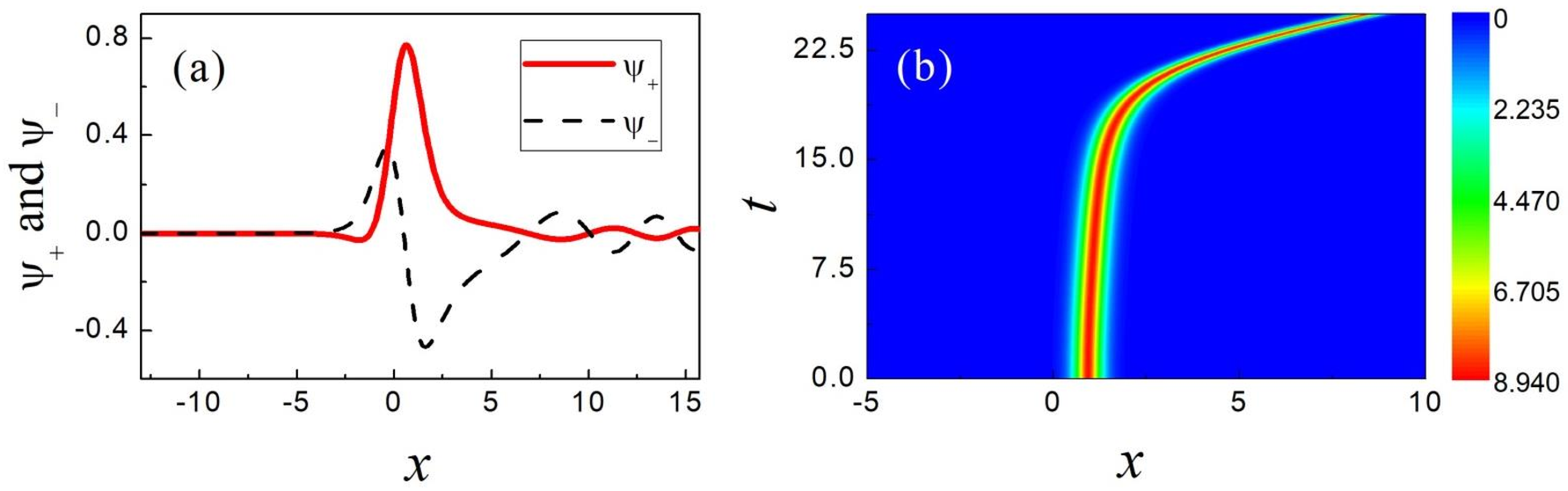}}
\caption{(a) A delocalized solution with nonvanishing tails, atr $%
(N,G,L)=(2,0.14,2)$. (b) The evolution of an unstable soliton, with
parameters $(N,G,L)=(6,0.28,2)$.}
\label{Tails}
\end{figure}

Results of the numerical analysis are summarized in Fig. \ref{Gcr}. First,
panel (a) displays a stability area for the stationary soliton solutions in
the $(N,G)$ plane. They exist and are stable in the yellow area, between $%
G=0 $ and the critical (largest) value $G=G_{\mathrm{cr}}$. Horizontal lines
connecting points $G=0$ and $G=G_{\mathrm{cr}}$ show that the variation of $%
G $ at fixed values of the chemical potential, $\mu $, practically does not
affect the respective value of the soliton's norm, $N$. At $G>G_{\mathrm{cr}%
} $, solutions with $N<3$ become delocalized, developing nonvanishing tails
which extend to the boundary of the integration domain, see Fig. \ref{Tails}%
(a). This result implies that, under the action of strong gravity, the
condensate leaks from the finite region at which the SO coupling is applied.
Further, in the interval of $3<N<5$, at $G>G_{\mathrm{cr}}$ the numerical
procedure yields only zero solution, while unstable solitons are obtained
for $N>5$ and $G-G_{\mathrm{cr}}$ positive but not too large. As shown in
Fig. \ref{Tails}(b), unstable solitons eventually escape from the trap and
start the free fall towards $x\rightarrow +\infty $. Only zero solution is
produced by the numerical method at any value of $N$ for $G-G_{\mathrm{cr}}$
large enough. In the case of $\gamma =1$ (the Manakov's nonlinearity),
values of $G_{\mathrm{cr}}$ for the SD and MM solitons are naturally found
to be identical. Typical examples of such stable solitons at the boundary of
their stability area are shown in Figs. \ref{StableSDMM}(a3,b3).

Figure \ref{Gcr}(a) clearly demonstrates that the stability area expands
with the increase of $N$, hence stronger nonlinearity makes the solitons
more stable against the action of the gravity. Additional numerical results
reveal that $G_{\mathrm{cr}}$ is smaller at both $\gamma <1$ and $\gamma >1$
than in the Manakov's case. In particular, Fig. \ref{Gcr}(a) shows the $G_{%
\mathrm{cr}}(N)$ curve for $\gamma =0$, with values $G_{\mathrm{cr}}$
essentially smaller than for $\gamma =1$. Thus, as mentioned above, the
Manakov's nonlinearity is optimal for the stabilization of the solitons
against gravity. Indeed, it is natural to expect that effects of the SO
coupling are strongest when the nonlinearity is equal for the nonlinear
self- and cross-interactions of components of the pseudo-spinor wave
function.

Further, Fig. \ref{Gcr}(b) shows the dependence of $G_{\mathrm{cr}}$ on size
$L$ of the confinement region in which the SO coupling is applied, as per
Eq. (\ref{Lambda}), for the SD soliton with $N=5$. The dependence features a
maximum at $L\approx 0.6$, which is, quite naturally, comparable to the
width of the corresponding soliton, $W\simeq $ $\sim 0.3$. The numerical
results demonstrate that $G_{\mathrm{cr}}$ decays as $L^{-1}$ with the
increase of $L$. This finding is explained below by means of the analytical
approximation, see Eq. (\ref{1/L}).

Finally, the growth of $G_{\mathrm{cr}}$ with the increase of the strength
of the localized SO-coupling term, i.e., $\Lambda _{0}$ in Eq. (\ref{Lambda}%
), is displayed in Fig. \ref{Gcr}(c). Roughly linear initial increase of $G_{%
\mathrm{cr}}$ with $\Lambda $, as well as its increase with $N$, observed in
the figure, also complies with Eq. (\ref{1/L}) derived below. The increase
of $\Lambda _{0}$ leads not only to the growth of critical gravity force, $%
G_{\mathrm{cr}}$, but also to growth of the amplitude of the nonzero tail of
the wave function extending towards the right edge of the solution domain at
$G>G_{\mathrm{cr}}$ (not shown here in detail).

\subsubsection{The analytical approximation}

Characteristics of the stationary state under the combined action of gravity
and SO-coupling-induced trap may be analyzed by means of the approximation
which treats the soliton as a quasi-particle. To illustrate the application
of the approximation in a simple form, in Appendix we formulate it for the
simplest setting based on the single GPE, with the localized trap
represented by an attractive delta-functional potential. Here we apply
similar analysis to the system under the consideration.

If the soliton's width is $\lesssim L$, then the effective soliton's
potential, induced by the localized SO coupling as per Eq. (\ref{H}), is
approximated as
\begin{equation}
U_{\mathrm{SOC}}(\xi )=-U_{0}\Lambda _{0}N^{2}\exp \left( -\xi
^{2}/L^{2}\right) ,  \label{U}
\end{equation}%
cf. Eq. (A5), where $\xi $\ is the coordinate of the soliton's center, $%
U_{0} $\ is a constant, and use is made of the fact that the soliton's width
and norm scale with its amplitude $A$ as $W\sim A^{-1}$ and $N\sim A^{2}W$ .
This potential gives rise to the trapping force,
\begin{equation}
F_{\mathrm{SOC}}(\xi )=-\frac{dU_{\mathrm{SOC}}(\xi )}{d\xi }=-2U_{0}\Lambda
_{0}N^{2}\left( \xi /L^{2}\right) \exp \left( -\xi ^{2}/L^{2}\right) .
\label{F}
\end{equation}%
The maximum value of the force is attained at the value of $\xi $\
determined by condition
\begin{equation}
\frac{dF_{\mathrm{SOC}}(\xi )}{d\xi }=0,~\mathrm{i.e.},~\xi _{\max
}^{2}=L^{2}/2,  \label{ximax}
\end{equation}%
and accordingly, the largest absolute value of the force corresponds to%
\begin{equation}
\left( F_{\mathrm{SOC}}(\xi )\right) _{\max }=-\sqrt{2/e}U_{0}\Lambda
_{0}N^{2}L^{-1}.  \label{Fmax}
\end{equation}

The competing gravity force acting on the soliton is estimated as%
\begin{equation}
F_{G}=F_{0}NG,  \label{FG}
\end{equation}%
where $F_{0}$\ is another constant. Finally, the largest value of $G$, up to
which the equilibrium condition, $F_{\mathrm{SOC}}(\xi )+F_{G}=0$, may hold,
is
\begin{equation}
G_{\mathrm{cr}}=\sqrt{\frac{2}{e}}\frac{U_{0}\Lambda _{0}}{F_{0}}\frac{N}{L},
\label{1/L}
\end{equation}%
cf. Eqs. (A6) and (A7) in appendix. Equation (\ref{1/L}) explains the
above-mentioned $L^{-1}$\ dependence for relatively large $L$. On the other
hand, if $L$ is too small, the SO-coupling trap is obviously weak, therefore
$G_{\mathrm{cr}}$ vanishes at $L\rightarrow 0$.

\subsection{Stationary states in the moving reference frame}

Stationary solutions in the moving reference frame were obtained by solving
Eq. (\ref{eq3}) with velocity $c\neq 0$. The procedure is similar to that
outlined above for $c=0$:

(i) For a given value of $c$, ITM is used to produce a stationary solution
with $G=0$, as the ground state.

(ii) Stable solitons with $G\neq 0$ are obtained by means of the
continuation of the latter solution by small steps to $G>0$, with the help
of SOM because, as well as in the case of $c=0$, the stationary soliton
cannot be a ground state at $G>0$.

\begin{figure}[h]
{\includegraphics[width=0.7\columnwidth]{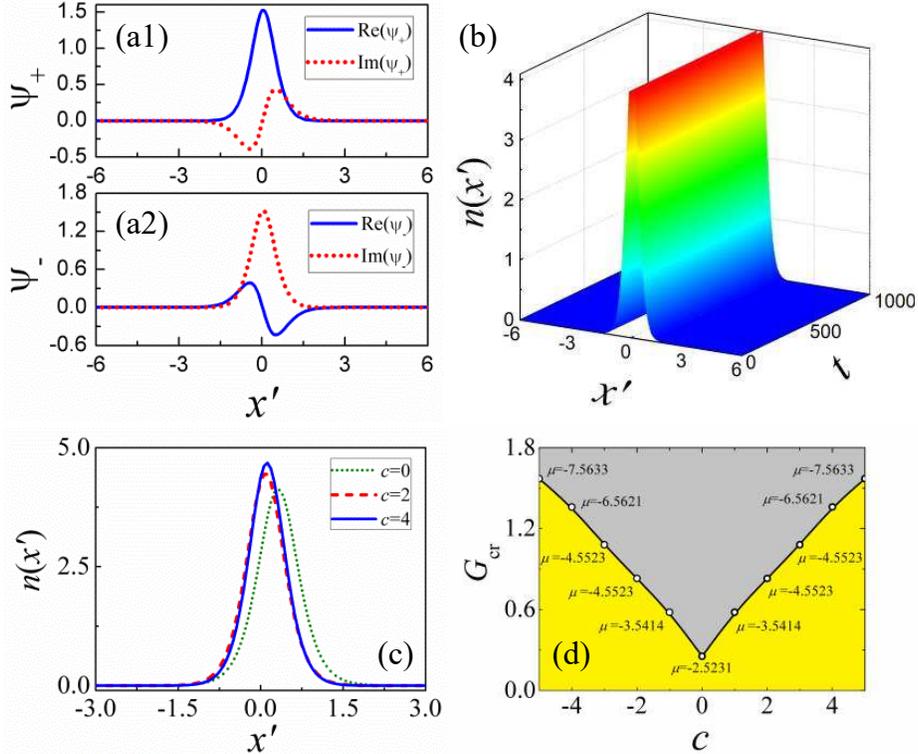}}
\caption{(a1,a2) Real and imaginary parts of two components of a composite
soliton of the SD\ type in the moving reference frame, with speed $c=4$, at $%
G=0.12$. (b) Direct simulations of the evolution of the soliton perturbed by
random noise at the $3\%$ amplitude level, which confirms its stability. The
evolution is shown by means of the spatiotemporal distribution of the total
density, $n(x^{\prime },t)$. (c) Total-density profiles of the stationary
solitons with a fixed value of the total norm, $N=4$, and different values
of velocity $c$. (d) The stability area (the yellow one), $G\leq G_{\mathrm{%
cr}} $, of the stationary solitons in the $(c,G)$ plane. In all the panels,
the parameters are fixed as $(N,L,\protect\gamma )=(4,2,1)$.}
\label{moving}
\end{figure}

An example of a stable soliton with $c\neq 0$ is displayed in Figs. \ref%
{moving}(a1,a2). The stability is corroborated by direct simulations of the
perturbed evolution, as shown in Fig. \ref{moving}(b). Comparison of the
field profiles displayed in Fig. \ref{moving}(a) with their counterparts,
shown in Fig. \ref{StableSDMM}(b2) for the same values $N=4$ and $G=0.12$,
but zero velocity, demonstrates that the moving soliton, remaining a stable
solution, develops a complex structure, in comparison with the essentially
real one at $c=0$. On the other hand, Fig. \ref{moving}(c) suggests that the
total-density profile of the moving soliton, shown in terms of $n(x^{\prime
})=|\phi _{+}(x^{\prime })|^{2}+|\phi _{-}(x^{\prime })|^{2}$, varies quite
slowly with the increase of the velocity at a fixed value of the total norm.
The latter figure demonstrates that the overall shape of the soliton
gradually becomes sharper, which enhances the strength of trapping by the
localized SO coupling.

Stability of the moving solitons of the SD type is summarized in the
parameter chart in the $\left( c,G\right) $ plane, which is displayed in
Fig. \ref{moving}(d). The symmetry of the stability area with respect to $%
c\longleftrightarrow -c$ corresponds to the above-mentioned invariance of
Eq. (\ref{eq3}) for the replacement of $c$ by $-c$. The figure demonstrates
that the increase of $|c|$ leads to \emph{additional stabilization} of the
trapped solitons (growth of $G_{\mathrm{cr}}$). This effect may be explained
by the fact that the linear-mixing terms $\sim ic\Lambda \left( x^{\prime
}\right) $ in Eq. (\ref{eq3}), being proportional to the spatial-modulation
coefficient, $\Lambda (x)$, induce an additional effective potential which
enhances the trapping action of the spatially confined SO coupling. In
particular, considering stationary solutions to Eq. (\ref{eq3}), $\psi _{\pm
}^{^{\prime }}=\exp \left( -i\mu t\right) \phi _{\pm }(x^{\prime })$, with $%
\left\vert \phi _{\_}\right\vert \ll \left\vert \phi _{+}\right\vert $, and
focusing on the effect of terms $\sim {ic\Lambda }\left( x^{\prime }\right) $
(neglecting, for the time being, terms $\sim d\Lambda (x^{\prime
})/dx^{\prime }$), the use of the second equation in system (\ref{eq3})
makes it possible to eliminate $\phi _{-}$ in favor of $\phi _{+}$, as $\phi
_{-}\approx -ic\mu ^{-1}\Lambda (x)\phi _{+}$. Then, the substitution of
this result in the first equation of system (\ref{eq3}) produces an
effective potential,%
\begin{equation}
U_{\mathrm{eff}}(x)\approx c^{2}\mu ^{-1}\Lambda ^{2}(x).  \label{Ueff}
\end{equation}%
Because soliton states exist at $\mu <0$, the potential given by Eq. (\ref%
{Ueff}) is indeed a trapping one, with the negative sign, thus helping to
additionally stabilize the bound state. The action of the extra potential
also explains the shift of the overall-density profile back towards $x=0$
and sharpening of the profile with the increase of $c$ in Fig. \ref{moving}%
(c).
\begin{figure}[h]
{\includegraphics[width=0.7\columnwidth]{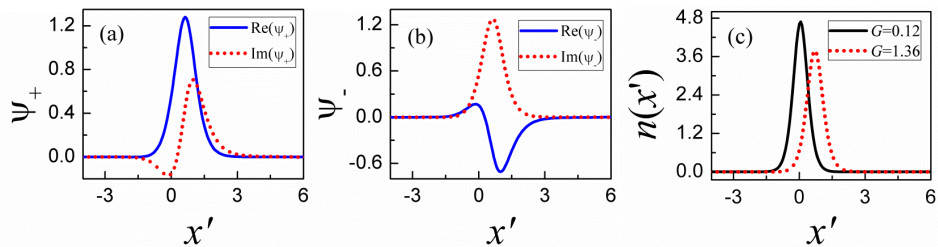}}
\caption{(a,b) Profiles of the two components of the moving soliton with the
same values $(N,L,\protect\gamma )=(4,2,1)$ as in Fig. \protect\ref{moving}%
(a1,a2,c), but much larger gravity force, $G=1.36$ (instead of $G=0.12$ in
Fig. \protect\ref{moving}). This soliton is located close to the stability
boundary, in terms of Fig. Fig. \protect\ref{moving}(d). (c) The comparison
of the total density of the same soliton (the red short dotted curves) and its
counterpart pertaining to $G=0.12$ (the solid black line).}
\label{former_Fig_5a}
\end{figure}

Lastly, while the moving solitons shown in Fig. \ref{moving}(a1,a2,c) are
located deep inside the stability area, in terms of Fig. \ref{moving}(d),
Fig. \ref{former_Fig_5a} displays an example of a soliton with the same
values of the norm, $N=4$, and speed, $c=4$, but taken at a much stronger
gravity force ($G=1.36$ instead of $G=0.12$), which places the soliton close
to the stability boundary. Comparison of the latter profile with its
counterpart from Fig. \ref{moving}(a1,a2), see panel (c) in Fig. \ref%
{former_Fig_5a}, provides another proof of the robustness of the solitons:
the increase of $G$ by a factor $\approx 11.3$ leads to a minor change in
the shape, \textit{viz}., reduction of the overall peak density by a factor $%
\approx 0.81$, and a shift of the peak towards $x>0$.

\section{Dynamical solutions}

\begin{figure}[h]
{\includegraphics[width=0.6\columnwidth]{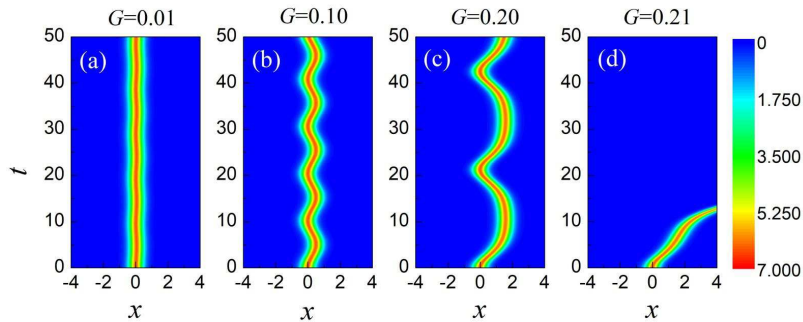}}
\caption{Direct simulation initiated by the abrupt application of gravity,
with strength $G=0.01$ (a), $0.10$ (b), $0.20$ (c), and $0.21$ (d), to a
soliton of the MM type, originally created for $G=0$. Other parameters are $%
(N,L,\protect\gamma )=(5,2,1)$. The evolution is shown by means of
spatiotemporal distributions of the total density.}
\label{oscillation}
\end{figure}
\begin{figure}[h]
{\includegraphics[width=0.4\columnwidth]{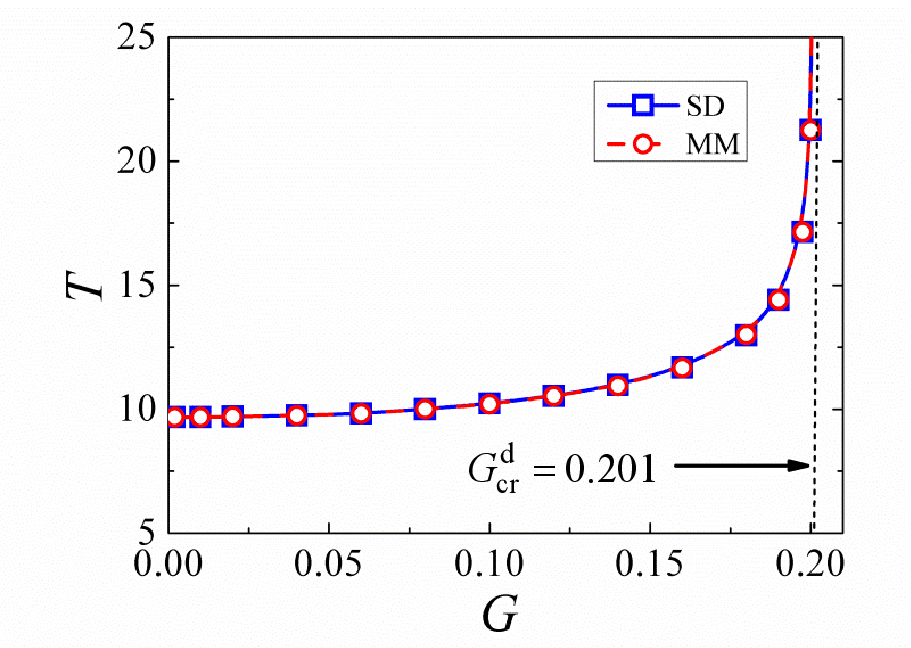}}
\caption{Period $T$ of the shuttle motion of solitons originally created
with $G=0$ and then abruptly set in motion by the application of gravity
force $G$. Dependences $T(G)$ are shown for solitons of the SD and MM types,
with parameters $(N,L,\protect\gamma )=(5,2,1)$. Note that the periods are
fully identical for both types of the solitons in the case of $\protect%
\gamma =1$ (the Manakov's nonlinearity). The soliton ecapes from the shuttle
regime at $G>G_{\mathrm{cr}}^{\mathrm{d}}$. }
\label{Period}
\end{figure}

The dynamics of solitons in the present system may be tested by abruptly
applying gravity to a mode created in a weightless environment, i.e.,
generated by Eq. (\ref{Basiceq}) with $G=0$ (in the experiment, this may be
realized by creating a soliton in a horizontally oriented cigar-shaped trap,
which is then quickly turned in the vertical plane). Figure \ref{oscillation}
exhibits a typical example of the subsequent evolution\ of a soliton of the
MM type, under the action of the suddenly applied gravity with different
values of $G$. It is seen that the soliton develops shuttle motion with an
amplitude and period depending on $G$ (oscillatory motion of nonlinear
SO-coupled wave packets in a trapping potential was considered in Ref. \cite%
{Mardonov}). The resulting period of the shuttle motion for the SD and MM
solitons is shown in Fig. \ref{Period} as a function of $G$. The figure
demonstrates that solitons of both types develop identical dynamics in the
case of the Manakov's nonlinearity, $\gamma =1$. The divergence of the
period at the critical value, $G=G_{\mathrm{cr}}^{\mathrm{d}}$, implies
that, at $G>G_{\mathrm{cr}}^{\mathrm{d}}$, the soliton can no longer be held
in the shuttle state by the spatially confined SO-coupling trap, see an
example in Fig. \ref{oscillation}(d)]. The dynamical critical value of the
gravity strength, $G_{\mathrm{cr}}^{\mathrm{d}}$, is presented, as a
function of $N$ and $L$, in Figs. \ref{Gcr}(a,b). Shapes of dependences $G_{%
\mathrm{cr}}^{\mathrm{d}}(N)$ and $G_{\mathrm{cr}}^{\mathrm{d}}(L)$ are
similar to their counterparts for the stationary solutions, while the
magnitude of $G_{\mathrm{cr}}^{\mathrm{d}}$ is, naturally, smaller, as
maintaining stable solitons in the dynamical regime requires to add some
margin to the critical value defined for the static state.

Long-time dynamics in the shuttle regime may be characterized by the
soliton's center-of-mass coordinate, defined as
\begin{equation}
X_{\mathrm{mc}}(t)={N}^{-1}\int_{-\infty }^{+\infty }(|\psi
_{+}(t)|^{2}+|\psi _{-}(t)|^{2})xdx.  \label{Xmc}
\end{equation}%
\begin{figure}[h]
{\includegraphics[width=0.5\columnwidth]{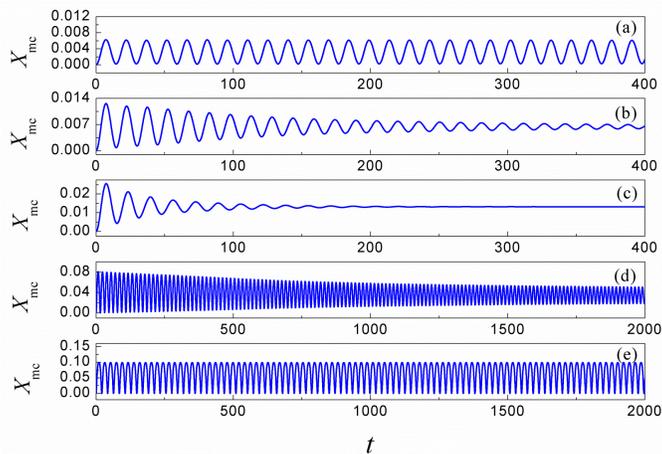}}
\caption{ (a,b,c) Motion of the center of mass of the MM soliton, originally
created with parameters $(N,\protect\gamma ,G_{0})=(1,1,0)$, under the
action of the suddenly applied gravity with $G=0.01$ (a), $0.02$ (b) and $%
0.04$ (c). (d) The same, but for initial parameters $(N,\protect\gamma %
,G_{0})=(3,1,0)$ and $G=0.16$. (e) The same for $(N,\protect\gamma %
,G_{0})=(5,1,0)$ and $G=0.2$. In all the cases, coefficients of the confined
SO-coupling term in Eq. (\protect\ref{Lambda}) are $\Lambda _{0}=1$ and $L=2$%
. }
\label{damped}
\end{figure}
Figure \ref{damped} displays the long-time regimes of motion produced by
simulations performed for different values of parameters $N$ and $G$. It is
observed that the oscillations are, generally, damped, which may be
explained by weak radiation losses of the soliton moving with acceleration
\cite{RMP}. The damping rate increases with the increase of $G$ and decrease
of $N$. The former peculiarity is explained by the fact that rapid
oscillations at smaller values of $G$ (see Fig. \ref{Period}) produce little
radiation \cite{RMP}, while, on the other hand, heavier solitons with a
larger norm are less amenable to the action of the weak recoil force
produced by the emission of radiation.

\section{Conclusion}

The objective of this work is to study the dynamics of two-component
matter-wave solitons under the action of the gravity field, with the aim to
hold the solitons in a fixed position, or move them at a constant speed, by
means of an effective trapping potential imposed by the spatially confined
SO (spin-orbit) coupling. In comparison with usual laser-beam tweezers,
advantage offered by the SO-coupling trap is the absence of detrimental
heating effects. First, stationary solutions of the system of coupled GPEs
for solitons of the SD (semi-discrete) and MM (mixed-mode) types were
produced, in the quiescent and moving reference frames. In both cases, the
critical value of the gravity force, $G_{\mathrm{cr}}$, which is the
boundary of the stability area for the solitons trapped by the localized SO
coupling, was found. By means of systematically collected numerical results
and the analytical approximation, which treats the soliton as a
quasi-particle, it was demonstrated that the stability area can be expanded
(i.e., $G_{\mathrm{cr}}$ made larger) by increasing the strength of the
SO-coupling term, the norm, and the speed, in the case of the moving system.
In the second part of the work, dynamical regimes were addressed, by
abruptly applying gravity to solitons created in the weightless setting. If
the gravity force is smaller than the respective dynamical critical value
(which, in turn, is smaller than its stationary counterpart), the solitons
feature damped shuttle motion, with the period and damping rate affected by
the gravity strength. The results produced by the consideration of both
stationary and dynamical settings demonstrate the ability of the effective
trapping potential, induced by the spatially confined SO coupling, to hold
solitons against the gravity, and transfer them at a constant speed, the
latter regime being actually more stable. These findings may help to design
new gravimeters and improve other techniques for precise measurements, based
on the use of cold atoms.

The present analysis may be extended in other directions. First, a natural
possibility is to consider this problem in the 2D geometry, where conditions
for the soliton stability are drastically different from those in the 1D
case \cite{Sakaguchi2014}-\cite{Pangw2018}. The 2D matter-wave solitons
stabilized by the SO coupling include vortex components, which suggests a
possibility to consider the interaction of vortex solitons with gravity.
Further, one can add the beyond-mean-field Lee-Huang-Yang corrections to the
GPE system \cite{Petrov}, and thus consider holding and transfer of
\textquotedblleft quantum droplets" (self-trapped modes with the flat-top
shape \cite{Petrov,recent-review}) in the presence of gravity. Note that the
interplay between the spatially uniform SO coupling and LHY corrections was
considered in Refs. \cite{Cui2018,QD2017,ZW2013}. A challenging option is to
extend the current setting to the full 3D geometry.

\begin{acknowledgments}
This work was supported, in part, by NNSFC (China) through Grant Nos.
11874112, 11905032, and 11805283, the Foundation for Distinguished Young
Talents in Higher Education of Guangdong through grant No. 2018KQNCX279, by
the Hunan Provincial Natural Science Foundation under Grant No. 2019JJ30044,
and by the Israel Science Foundation (project No. 1287/17).
\end{acknowledgments}

\section*{Appendix: The analytical approximation in the simplest form}

The single GPE, which includes the gravity field and the localized trap
represented by the $\delta $-functional attractive potential with strength $%
\varepsilon >0$ (such as one induced by a tightly focused red-shifted laser
beam), is%
\begin{equation}
i\partial _{t}\psi =-{\frac{1}{2}}\partial _{xx}\psi -|\psi |^{2}\psi
-Gx\psi -\varepsilon \delta (x)\psi .  \tag{A1}
\end{equation}%
Here the localized potential is set at $x=0$ without the loss of generality.
The Hamiltonian corresponding to Eq. (A1) is%
\begin{equation}
H=\int_{-\infty }^{+\infty }\left[ \frac{1}{2}\left( \left\vert \partial
_{x}\psi \right\vert ^{2}-\left\vert \psi \right\vert ^{4}\right)
-Gx\left\vert \psi \right\vert ^{2}\right] dx-\varepsilon |\psi (x=0)|^{2},
\tag{A2}
\end{equation}%
cf. Eq. (\ref{H}). Stationary solutions to Eq. (A1), with chemical potential
$\mu $, are looked form as $\psi =\exp \left( -i\mu t\right) \phi (x)$,
where real function $\phi (x)$ obeys equation%
\begin{equation}
\mu \phi =-{\frac{1}{2}}\frac{d^{2}\phi }{dx^{2}}-\phi ^{3}-Gx\phi
-\varepsilon \delta (x)\phi .  \tag{A3}
\end{equation}

In the absence of the gravity and trapping potential, the commonly known
nonlinear-Schr\"{o}dinger soliton with norm $N$, $\mu =-N^{2}/8$, and
central coordinate $\xi $, is
\begin{equation}
\phi _{\mathrm{sol}}(x)=(N/2)\mathrm{sech}\left( (N/2)\left( x-\xi \right)
\right) .  \tag{A4}
\end{equation}%
If the gravitational and trapping potentials in Eq. (A2) are considered as
small perturbations, the respective terms in the soliton's energy are%
\begin{equation}
U_{\varepsilon }(\xi )+U_{G}(\xi )=-\left( N/2\right) ^{2}\varepsilon ~%
\mathrm{sech}^{2}\left( N\xi /2\right) -GN\xi ,  \tag{A5}
\end{equation}%
which predicts a stationary state at a local minimum of the potential,
defined by $\left( d/d\xi \right) \left( U_{G}(\xi )+U_{\varepsilon }(\xi
)\right) =0$, i.e., at $\xi $ which is a smaller root of the equation
\begin{equation}
\tanh \left( \frac{N}{2}\xi \right) -\tanh ^{3}\left( \frac{N}{2}\xi \right)
=\frac{4G}{\varepsilon N^{2}}.  \tag{A6}
\end{equation}%
As follows from Eq. (A6), the stationary state exists at values of the
gravity force%
\begin{equation}
G\leq G_{\mathrm{cr}}=\frac{\varepsilon N^{2}}{6\sqrt{3}}.  \tag{A7}
\end{equation}%
Alternatively, one can conclude that, for given $G$, the trapping potential
is able to hold solitons with norms exceeding a certain minimum value,%
\begin{equation}
N\geq N_{\min }=\sqrt{6\sqrt{3}G/\varepsilon }.  \tag{A8}
\end{equation}

\end{document}